\documentclass[aps,pra,twocolumn,superscriptaddress,nofootinbib]{revtex4-2}

\usepackage{xcolor}
\usepackage{multirow}
\usepackage{array}
\usepackage{makecell}
\usepackage{braket}
\usepackage{graphicx}
\usepackage{amsmath}
\usepackage{amssymb}
\usepackage{hyperref}
\usepackage{enumitem}
\usepackage{float}

\setcellgapes{3pt}
\makegapedcells

\begin{document}

\title{Schr\"odinger's unpublished hidden-variable model in the 1944--45 \\ debate over the ``Collapse of Determinism''}

\author{Flavio Del Santo}
\affiliation{University of Geneva, Group of Applied Physics, Rue de l'École-de-Médecine 21, Geneva 1211, Switzerland}
\affiliation{Constructor University, Bremen, Germany}

\begin{abstract}
\noindent In the mid-1940s, the question of whether quantum mechanics could be supplemented by deterministic hidden variables resurfaced in a debate  in \textit{Nature}. I show that  Erwin Schr\"odinger drafted a hitherto unpublished response to this controversy. In it, Schr\"odinger provided an explicit hidden-variable model, thereby showing that such models could not be ruled out. I reconstruct the historical context of this forgotten manuscript and publish here the first transcription and analysis. Importantly, Schr\"odinger's proposal should be understood as a proof of principle rather than as evidence of  his commitment to determinism.
\end{abstract}
\maketitle
\section{Are hidden variables (im)possible?}
\label{intro}

When we toss a fair coin, we say that there is a fifty percent chance that it will land, say, heads. According to classical physics, however, this apparent randomness merely reflects our ignorance of all the relevant parameters of the toss, such as the exact magnitude and direction of the applied force, the air resistance, etc. Were we able to specify all these variables with absolute precision, we would find that the behavior of the coin is, in fact, deterministic.\footnote{Note that this standard view of classical determinism was questioned by several authors, including Erwin Schr\"odinger \cite{Schrodinger1932Indeterminismus}, Karl Popper \cite{Popper1950Indeterminism}, Max Born \cite{born1969classical}, and others \cite{del2019physics}.}

Things changed with the advent of quantum physics. The quantum formalism itself generally allows us to predict only probabilities for the outcomes of measurements. Since the early days of the theory, 
several physicists started wondering whether it would be possible to restore determinism by assuming that the quantum-mechanical description was incomplete and could be supplemented by \textit{hidden variables}, much as in the classical description of a coin toss \cite{bubhv}. Early attempts in this direction include Louis de Broglie's pilot-wave theory \cite{deBroglie1927mecanique}, as well as a hidden-variable model devised by Albert Einstein  (which he never published)  \cite{belousek1996einstein}.

In 1932, however, John von Neumann, one of the foremost authorities on the mathematical foundations of quantum mechanics, presented what was widely taken to be a proof that quantum theory could not be supplemented by additional hidden variables \cite{vonNeumann1932}. Although Grete Hermann, as early as 1935, identified issues with the generality of von Neumann's argument, her criticism went almost entirely unnoticed, possibly also because she was ``young, a woman and an outsider to the discussion among physicists'' \cite{crull2016grete}. The limitations of von Neumann's argument would become widely recognized much later, after David Bohm's 1952 complete deterministic hidden-variable interpretation  \cite{bohm1952suggested}, and John Bell's criticism --- closely related to Hermann's --- in 1966  \cite{bell1966problem,mermin2018homer}.

Nevertheless, in the meantime, the search for hidden-variable models had essentially disappeared for over a decade, certainly also because of von Neumann's result and his overwhelming authority.\footnote{The most notable exception was the celebrated Einstein-Podolsky-Rosen paper \cite{einstein1935can}, which argued for the incompleteness of the quantum formalism, without however constructing a hidden-variable model.}

It was not until the mid-1940s that the debate over classical and quantum indeterminism suddenly resurfaced, despite foundational discussions of quantum mechanics having been stifled by the war that was shaking the world. This revival was sparked by two independent events: the first was the twenty-seventh Guthrie Lecture, delivered on May 18, 1943, at the Physical Society in London by the eminent British mathematician and physicist Edmund Taylor Whittaker, under the title ``Chance, Freewill and Necessity in the Scientific Conception of the Universe'' \cite{Whittaker_1943}. The second was the presidential address ``The Classical Determinism and the Quantum Theory'', delivered by the great Indian physicist Satyendra Nath Bose at the Indian Science Congress in Delhi on January 3, 1944   \cite{bose1944classical}.

Both contributions independently put forward roughly the same fundamental thesis: quantum mechanics had provided enough grounds to finally question once and for all the classical dogma of determinism in physics. This is vividly expressed by Bose when he states: ``Fifty years ago the belief in causality and determination was absolute. To-day physicists have gained knowledge but lost their faith" \cite{bose1944classical}. 

What is crucial to note is that, on the one hand, both authors agree that there is nothing mysterious about classical statistical results, since, in the words of Whittaker, these are ``\textit{crypto-deterministic}: by which is meant that the phenomenon is in reality deterministic, although we cannot foretell its outcome on account of our lack of information regarding certain hidden parameters'' \cite{Whittaker_1943}. On the other hand, they both go on to propose a number of arguments aimed at showing that such a hidden-variables explanation is fundamentally impossible in quantum mechanics, also explicitly appealing in both cases to von Neumann's proof. As we will see, while both Whittaker and Bose reinforced von Neumann's case against hidden variables, their contributions also reignited a debate that ultimately renewed skepticism about the supposed impossibility of hidden-variable theories.

Despite not taking a categorical position, and instead presenting determinism and indeterminism as two opposing camps, Bose's cogent words clearly betray his sympathy for the latter:
``Causality and the universal laws are to be thrown simultaneously overboard. These assertions are so revolutionary that, no wonder, they have forced physicists to opposing camps. There are some who look upon causality as an indispensable postulate for all scientific activities. [...] Others of the opposite camp look upon old determinism as an inhuman conception, not only because it sets up an impossible ideal, but also as it forces man to a fatalistic attitude which regards humanity as inanimate automata in the hands of an iron law of causation'' \cite{bose1944classical}.

In his contribution, Whittaker shows an even stronger commitment to irreducible indeterminism, on the grounds that quantum mechanics is supposedly incompatible with additional hidden variables: ``von Neumann has shown that `chance', as manifested in quantum
phenomena, cannot be accounted for by postulating hidden parameters, but
represents a true indeterminacy. The bonds of necessity have been broken;
for certain classes of phenomena, crypto-determinism is definitely disproved" \cite{Whittaker_1943}.
And besides appealing to ``the more elaborate mathematical proof" of von Neumann, Whittaker even puts forward a more intuitive physical argument: he purports to show that no deterministic hidden-variable model could account for the quantum Malus' law. This asserts that a single linearly polarized photon is transmitted through a polarizer with probability equal to $cos^2 \varphi$, where $\varphi$ is the angle between the photon's polarization and the transmission axis of the polarizer.\footnote{Whittaker and the subsequent contributors refer to the polarizer as a ``Nicol prism'', a calcite device named after its inventor, William Nicol, and widely used at the time.}

The argument (slightly adapted into modern terms) goes as follows: Suppose that a photon carries, as a hidden variable $\lambda$, one bit of information whose two possible values uniquely determine whether the photon will be transmitted (say, for $\lambda=0$) or reflected (for $\lambda=1$). This is much like a classical coin that, after the toss, carries one bit of information whose value --- hidden from us merely by our ignorance --- uniquely determines whether the outcome is ``heads'' or ``tails''. For an ensemble of $N$ photons to reproduce the correct quantum statistics for a given polarizer set at an angle $\varphi$, $N\cos^2\varphi$ photons must therefore carry the value $\lambda=0$, while $N\sin^2\varphi$ must carry the value $\lambda=1$. But once the ensemble has been prepared and the photons are already in flight toward the polarizer, one may arbitrarily change the latter's orientation. Changing the orientation of the polarizer would not change the values of $\lambda$ already carried by the photons, and hence would leave the predicted transmission fraction unchanged. Yet Malus' law requires this fraction to vary with $\varphi$. Whittaker thus concludes that a hidden-variable description is impossible.

Likely because of its simplicity compared with von Neumann's abstract result, it was mainly this latter example that was to trigger a whole debate which unfolded in the pages of the journal \textit{Nature}. The debate went on to involve several scholars, who published eleven contributions over the course of around a year (Sect. \ref{debate}).\footnote{Ten of these appeared as contributions in \textit{Nature} \cite{naturePiaggio1944Collapse, natureWest1944Collapse, naturePiaggio1944CollapseReply, natureBarkas1944Collapse, natureWhittaker1944Collapse, natureGold1945Causality, natureHorzelski1945Causality, natureWest1945Causality, naturePiaggio1945Causality, natureBrabazon1945Causality} and one in the \textit{Proceedings of the Physical Society} \cite{pelzer1944interpretation}.} What had not been known until now, however, is that there was also a twelfth contribution, which never appeared in print and was authored by no less a figure than one of the founding fathers of quantum mechanics, Erwin Schr\"odinger \cite{schrodinger1944collapse}. In his reply to this exchange, Schr\"odinger provided an explicit hidden-variable model that disproved Whittaker's argument. I present here the first full transcription (Appendix~\ref{appendix}) and a detailed analysis of Schr\"odinger's manuscript (Sect. \ref{hvmodel}).

\section{The debate in the pages of \textit{Nature}}
\label{debate}

The aforementioned talks by Whittaker and Bose --- and, with them, the common question of whether quantum mechanics could be supplemented by hidden variables --- both found their way into the prestigious journal \textit{Nature}: the former as a short anonymous news article \cite{natureChanceFreewillNecessity1943}; the latter through an excellent and extensive report written by the British mathematician Henry T. H. Piaggio, entitled ``Collapse of Determinism'' \cite{naturePiaggio1944Collapse}, which proved to be the real trigger of the debate.

The first reactions were not long in coming. In October 1944, \textit{Nature} published a communication by experimentalist Gilbert D. West in reply to Piaggio's article and under the same heading \cite{natureWest1944Collapse}. There, he objects to Piaggio's ``claim that the existence of causality is disproved'' \cite{naturePiaggio1944Collapse}, noting that the same claim had been a central theme of Whittaker's Guthrie lecture, but remained, in his view, open to severe criticism. In particular, he questions the straightforward application of von Neumann's mathematical proof to physical reality: ``there are many who, knowing how easily assumptions can lurk in analytical work, refuse to be mathematically bludgeoned. Harm can be done to the reputation of science by any who proclaim as a definite doctrine what is but a tentative effort" \cite{natureWest1944Collapse}.

Piaggio's response is published alongside West's letter. Therein, he claims that his view is, in fact, very similar to West's and that, writing on a controversial subject, he found it ``necessary to mention all interpretations given by eminent authorities" \cite{naturePiaggio1944CollapseReply}. He concludes with a similar warning of caution towards the interpretation of mathematical results: ``I wish to put forward a general argument against all supposed demonstrations of impossibility, such as von Neumann's. However careful the logic, at best all that has been demonstrated is that the impossibility holds for the problem as formulated" \cite{naturePiaggio1944CollapseReply}.

The debate continued the following month with a letter from Wilfred W. Barkas, who raises the question: ``how could individually indeterminate photon outcomes nevertheless give rise to regular statistical results? \cite{natureBarkas1944Collapse} 
Whittaker's reply, published alongside Barkas's letter, appeals to repeated coin tosses: he notes that their statistical regularity follows from the ordinary theory of probability together with the symmetry of the individual trial, without any need to assume a crypto-deterministic mechanism \cite{natureWhittaker1944Collapse}.

The \textit{Nature} debate continued in January 1945 with a series of three letters appearing under the new heading ``Causality or Indeterminism?'' \cite{natureGold1945Causality,natureHorzelski1945Causality,natureWest1945Causality}. There, the prominent British meteorologist Ernest Gold questioned Whittaker's appeal to symmetry in the classical coin-tossing example, pointing out that the mechanism producing the toss need not itself behave randomly \cite{natureGold1945Causality}. Jerzy Horzelski, in turn, reports on a simple experiment of his own, using a mechanical device to keep the conditions of the toss approximately fixed. He found the outcome to be overwhelmingly reproducible under the same conditions, thus empirically reinforcing the classical crypto-deterministic picture \cite{natureHorzelski1945Causality}.

The most significant step, however, came in a further contribution by West \cite{natureWest1945Causality}. He observes that Whittaker's argument could only establish the failure of a causal explanation in which the relevant parameters are assumed to belong to the photon alone. As he puts it: ``What, under certain conditions, [Whittaker's argument] does show is that causality fails if parameters are restricted to photons. Clearly, however, in the passage of photons through Iceland spar [i.e., the polarizer], parameters associated with the crystal lattice must be taken into account, and when this is done, the argument fails'' \cite{natureWest1945Causality}. We will see in the next section that this indeed was an important intuition. West then adds: ``Apparently von Neumann's argument has a similar weakness'' \cite{natureWest1945Causality}, citing in this connection a paper by the Austrian-born physicist Heinrich Pelzer. Pelzer had not, in fact, proved von Neumann wrong, but had proposed in 1944 rudimentary hidden-variable models showing that Whittaker's argument based on Malus' law had not established the impossibility of a deterministic description \cite{pelzer1944interpretation}. We  will return to Pelzer's result and its possible connection with Schr\"odinger's unpublished manuscript in Sec. \ref{whynot}.

The debate in \textit{Nature} essentially came to an end in March 1945 with another article by Piaggio,\footnote{With the exception of another, not really relevant letter \cite{natureBrabazon1945Causality}.} prompted by the fact that the topic had ``produced further letters, too numerous for the Editors to publish in full'', so that he had been invited to give a ``connected account of the points raised'' \cite{naturePiaggio1945Causality}. Piaggio's paper offers a remarkably clear synthesis of the debate, while adding new elements and references. In particular, he presents von Neumann's argument in greater detail, discusses Pelzer's counterexamples \cite{pelzer1944interpretation}, and puts forward his own doubts about von Neumann's proof on the basis of the Einstein-Podolsky-Rosen argument \cite{einstein1935can}.

In Piaggio's account, we can say that the debate ends with a truce: ``I therefore, with great diffidence, offer the opinion that the existence of causality has \textit{not} been disproved. [...] I will quote the opinion expressed by Bertrand Russell in 1936, that at present there is no decisive reason in favor of complete determinism (causality) in physics, but there is no reason against it'' \cite{naturePiaggio1945Causality}.

\section{Schr\"odinger's unpublished manuscript}
\label{hvmodel}
Schr\"odinger's response to the debate survives as an undated handwritten manuscript entitled ``Collapse of Determinism'' \cite{schrodinger1944collapse}, which never appeared in print. In the manuscript, there is explicit mention of the \textit{Nature} debate up to Barkas's letter of November 25, 1944 \cite{natureBarkas1944Collapse}, thereby providing a \textit{terminus post quem} for its drafting. A full transcription is provided in Appendix~\ref{appendix}. In this section, I analyze the content of the manuscript in detail, first reconstructing Schr\"odinger's hidden-variable model and then discussing the broader philosophical considerations present in the conclusions of the manuscript.\footnote{I will occasionally change the notation (e.g., using $\lambda$ for the hidden variable instead of the original $x$ and ``polarizer'' instead of ``Nicol''), and expand on Schr\"odinger's original explanation where further clarification is needed. For historical accuracy, the reader should always refer to the original manuscript reproduced in Appendix~\ref{appendix}.}

\subsection{Schr\"odinger's deterministic hidden-variable model of Malus' law}

After restating Whittaker's argument \cite{Whittaker_1943}, which purported to establish the impossibility of a hidden-variable explanation of Malus' law (see Sect. \ref{intro}), Schr\"odinger declares: ``I too have been puzzled by this conclusion. For the derived absurdity is visibly produced by the simplifying assumption of two discrete parameter values only."

He then presents the following model. Assume that each photon carries instead a hidden variable given by a real number $\lambda\in[0,1]$, uniformly distributed, with $\rho(\lambda)=1$. Hence, out of the $N$ photons in the initial ensemble, $N\,d\lambda$ have a value of the hidden variable in the interval $[\lambda,\lambda+d\lambda]$.

The photons then travel towards a polarizer oriented at an angle $\varphi$, as in Whittaker's original scenario. Schr\"odinger, however, posits the following deterministic rule for transmission ($T$), which is a function of both the hidden variable and the setting of the polarizer:
\begin{equation}
T(\lambda,\varphi)=
\begin{cases}
1, & \lambda\geq \sin^2\varphi,\\
0, & \text{otherwise}.
\end{cases}
\end{equation}

The number of transmitted photons is therefore
\begin{equation}
N_T
=
N \int_{\sin^2\varphi}^{1}\rho(\lambda)\,d\lambda
=
N \cos^2\varphi,
\end{equation}
so that Malus' law is recovered.\footnote{Schr\"odinger treats explicitly also the reflection which we omit here because the argument proceeds analogously.}

Schr\"odinger realized, however, that a difficulty would arise with subsequent measurements. After transmission through the first polarizer, only photons
satisfying $\lambda\geq\sin^2\varphi$ remain in the ensemble. The distribution of
the hidden variable is therefore no longer uniform on $[0,1]$, but becomes (not explicitly written by Schr\"odinger), $\rho(\lambda\mid T)=1/\cos^2\varphi$, for $\sin^2\varphi\leq\lambda\leq1$, and 0 otherwise.

The model therefore requires a post-transmission update rule for the hidden variable. The rule must be deterministic in order for the whole argument against Whittaker to hold. Schr\"odinger provides exactly such an update rule, which is a function of $\lambda$ and $\varphi$:
\begin{equation}
\lambda'
=
\frac{\lambda-\sin^2\varphi}{\cos^2\varphi}.
\end{equation}
This maps the interval $[\sin^2\varphi,1]$ onto $[0,1]$, so that the updated hidden variable $\lambda'$ is
again uniformly distributed.
Schr\"odinger's explicit construction ends here, but it is instructive to show why the update rule allows one to reproduce the quantum prediction for a second polarizer.

Consider now a second polarizer oriented at an angle $\varphi'$. Since all photons
emerging from the first polarizer are polarized along $\varphi$, the relevant
relative angle is $\varphi'-\varphi$. The transmission rule for the second polarizer
is therefore
\begin{equation}
T'(\lambda',\varphi, \varphi')=
\begin{cases}
1, & \lambda'\geq \sin^2(\varphi'-\varphi),\\
0, & \text{otherwise}.
\end{cases}
\end{equation}

Hence
\begin{equation}
N_{T'}
=
N_T\int_{\sin^2(\varphi'-\varphi)}^{1} d\lambda'
=
N_T\cos^2(\varphi'-\varphi),
\end{equation}
in agreement with the  prediction of quantum mechanics, which completes the argument. 

A point that appears to have escaped all the participants in the debate, including both Whittaker and Schr\"odinger, is that there is nothing specifically quantum about the inadequacy of a one-bit hidden variable. If all the parameters of an experiment are fixed and the outcome is binary, one bit is sufficient to encode the predetermined outcome of that fixed experiment, either in classical or quantum physics. In Whittaker's example, however, the apparatus itself contains a continuously adjustable parameter, namely the orientation $\varphi$ of the polarizer. One bit is therefore no longer sufficient: the outcome must depend on the interplay between a continuous  hidden variable and the setting of the apparatus, as in Schr\"odinger's rule for $T(\lambda,\varphi)$.

Yet the same situation can arise with a classical coin. If the height of the floor is fixed, one bit can encode whether the coin will land heads or tails. But if, for instance, the floor can be moved while the coin is still in flight, the outcome depends on both the state of the coin and the continuously variable height of the floor. Thus, the need for a real-number hidden variable in Schr\"odinger's construction has nothing specifically quantum about it.
\subsection{Philosophical considerations}

The remainder of the manuscript is equally interesting, as Schr\"odinger turns from the specific model to the broader status of determinism and indeterminism. He writes:
``even very elaborate
and mathematically rigorous deductions, such as Von Neumann's, are of limited
value. Experimental evidence + logical deduction can \underline{never} prove that either
the deterministic or the indeterministic view is \underline{impossible}. At the very best
the sum-total of evidence can lead to the conclusion that one or the other is
[...] extremely
inconvenient, because it introduces unwarrantable hypotheses."

Moreover, Schrödinger states explicitly that, even if the issue cannot be decided on purely logical or empirical grounds, he believes that “the strongest plea in favour of indeterminism is older than quantum mechanics.” He explains that “we know to day [\textit{sic}] that our idea of a lawful sequence of events in Nature has been abstracted from events which are based directly on statistical law.” Nevertheless, because the idea of a ``law of Nature'' had become deeply rooted before the statistical character of those laws was appreciated, ``we are tempted to construct in our mind `true microscopic lawfulness' behind the observed statistical one.''

This conclusion is particularly important for interpreting the status of Schr\"odinger's hidden-variable model. Far from signaling a conversion to determinism, it is fully consistent with a commitment to indeterminism that he had developed long before, and independently of, quantum mechanics \cite{Schrodinger1932Indeterminismus}. Already in the early 1920s Schr\"odinger had embraced the view, inherited especially from Franz Exner, that ``Nature's laws are of thoroughly statistical character'', maintained that ``classical mechanics itself is indeterminate'', and criticized determinism as the ``fundamental dogma of classical physics'', a ``philosophic prejudice'', and an ``a priori belief'' \cite{del2026schrodinger}. Schr\"odinger's model should therefore be understood not as an endorsement of a deterministic picture of nature, but as a proof of principle: determinism might not be necessary or even particularly plausible, but neither could it be ruled out by an alleged proof of logical impossibility.

\section{Why did Schr\"odinger not publish his model?}
\label{whynot}

As we have previously mentioned, Pelzer had already published a paper (communicated on January 21, 1944, and published on May 1) presenting a hidden-variable explanation of Whittaker's example \cite{pelzer1944interpretation}. Pelzer's paper is, however, rather unusual in form. He first provides two ``admittedly rather artificial'' mechanical analogues, with basically no mathematical formalization. In the first, photons are represented as thin needle-like rods, while the polarizer is modeled by two parallel gratings, the second oscillating irregularly with respect to the first and reorienting the transmitted rods. In the second, photons are represented by material balls of different diameters, due to different temperatures, which pass through a sieve only if their diameter lies below a certain threshold; a subsequent heating mechanism then reshuffles their distribution.

Only in a very short ``\textit{Note added in proof}'' does Pelzer provide an explicit quantitative hidden-variable model. Each photon is assigned a continuous parameter $\lambda$, with $0<\lambda<\pi/2$, and is transmitted through a polarizer tilted by an angle $\varphi$ whenever $\lambda>\varphi$. By assuming a distribution proportional to $\sin 2\lambda\,d\lambda$, Pelzer recovers the $\cos^2\varphi$ transmission probability. At the level of a single polarizer, this model is mathematically equivalent to Schr\"odinger's under the simple reparameterization $x=\sin^2\lambda$.

Interestingly, Pelzer acknowledges von Neumann for having pointed out the crucial aspect of considering multiple successive polarizers. To account for these, Pelzer states that, after transmission, the hidden variables need to be reshuffled so as to restore the same distribution. Unlike Schr\"odinger, however, he does not specify an explicit deterministic transformation law for the hidden variable of each individual photon.

It is plausible that Schr\"odinger became aware of Pelzer's paper through its citation in the \textit{Nature} debate, in particular in West's communication of January 1945 \cite{natureWest1945Causality}. Schr\"odinger's draft shows that he had followed the debate at least up to the contributions published on November 25, 1944. If he subsequently learned of Pelzer's result, he may have regarded the essential point as already established and therefore decided not to pursue publication of his own reply. Schr\"odinger's manuscript was nevertheless considerably broader in scope and philosophically more ambitious, combining a concrete hidden-variable model with a reflection on the limits of impossibility proofs and on the respective status of deterministic and indeterministic descriptions.

    
 \section{Conclusion}

I have here reconstructed a largely forgotten debate that, after more than a decade in which the constructive search for hidden-variable models had disappeared, reopened between 1943 and 1945 the question of whether quantum mechanics could admit a deterministic completion. A central new result is that Schr\"odinger himself conceived a hidden-variable model in the context of this debate, although he never published it. 

One would have to wait until 1952 for hidden variables to re-emerge forcefully in the debate on the foundations of physics, with David Bohm's \textit{causal interpretation} of quantum mechanics \cite{bohm1952suggested}. A further turning point came in 1964, when Bell showed that local hidden-variable models lead to experimentally testable predictions \cite{bell1964einstein}, and subsequent experiments have by now ruled out this class of models \cite{kaiser2020tackling}.

The debate between determinism and indeterminism has divided physicists for over a century and continues to do so today. From the outset, it has carried a strong ideological charge, often taking the form of a partisan confrontation between two opposing camps, each electing its own champions and, at times, instrumentalizing them in support of positions that ran against their own views.

The paradigmatic case is that of Einstein, Schr\"odinger, and Bohm, perhaps the three figures most commonly regarded, in the popular culture of physicists, as the main authorities on the side of determinism. Indeed, all three devised, at different moments, hidden-variable models, yet none of them can simply be characterized as an advocate of determinism.
As a matter of fact, Einstein remained uneasy about proposals for a deterministic completion of quantum mechanics \cite{DelSanto2019Realism}. Bohm was openly critical of mechanistic determinism and wrote extensively against it \cite{DelSantoKrizek2025Bohm,VanStrien2025Bohm}. Schr\"odinger, finally, had developed a remarkably strong commitment to indeterminism even before, and independently of, quantum mechanics, which he maintained throughout his career \cite{hanle1979indeterminacy, stoltzner1999vienna,Stoeltzner2012Schrodinger,del2026schrodinger}.

What they have in common is that they devoted much of their intellectual lives to questions of causality and indeterminism, which led them, at times, to explore deterministic models and their limits. Such models could serve as proofs of principle against widespread beliefs or alleged impossibility claims, as strikingly illustrated by the case of Schr\"odinger reconstructed here, but never as expressions of a dogmatic commitment to determinism.

\clearpage
\bibliographystyle{unsrt}
\bibliography{bibliography}

\begin{thebibliography}{10}

\bibitem{Schrodinger1932Indeterminismus}
Erwin Schr{\"o}dinger.
\newblock {\"U}ber indeterminismus in der physik.
\newblock In {\em Zwei Vortr{\"a}ge zur Kritik der naturwissenschaftlichen Erkenntnis}, pages 1--24. Johann Ambrosius Barth, Leipzig, 1932.

\bibitem{Popper1950Indeterminism}
Karl~R. Popper.
\newblock Indeterminism in quantum physics and in classical physics.
\newblock {\em The British Journal for the Philosophy of Science}, 1(2):117--133, August 1950.

\bibitem{born1969classical}
Max Born.
\newblock Is classical mechanics in fact deterministic?
\newblock In {\em Physics in my Generation}, pages 78--83. Springer, 1969.

\bibitem{del2019physics}
Flavio Del~Santo and Nicolas Gisin.
\newblock Physics without determinism: Alternative interpretations of classical physics.
\newblock {\em Physical Review A}, 100(6):062107, 2019.

\bibitem{bubhv}
Jeffrey Bub.
\newblock Hidden variables.
\newblock In Olival Freire, editor, {\em The Oxford Handbook of the History of Quantum Interpretations}. Oxford University Press, 03 2022.

\bibitem{deBroglie1927mecanique}
Louis de~Broglie.
\newblock La m{\'e}canique ondulatoire et la structure atomique de la mati{\`e}re et du rayonnement.
\newblock {\em Journal de Physique et le Radium}, 8(5):225--241, 1927.

\bibitem{belousek1996einstein}
Darrin~W Belousek.
\newblock {Einstein}'s 1927 unpublished hidden-variable theory: Its background, context and significance.
\newblock {\em Studies in History and Philosophy of Modern Physics}, 27(4):437--461, 1996.

\bibitem{vonNeumann1932}
John von Neumann.
\newblock {\em Mathematische Grundlagen der Quantenmechanik}, volume~38 of {\em Grundlehren der mathematischen Wissenschaften}.
\newblock Julius Springer, Berlin, 1932.

\bibitem{crull2016grete}
Elise Crull and Guido Bacciagaluppi.
\newblock {\em {Grete Hermann}-Between Physics and Philosophy}.
\newblock Springer, 2016.

\bibitem{bohm1952suggested}
David Bohm.
\newblock A suggested interpretation of the quantum theory in terms of ``hidden'' variables, parts {I} and {II}.
\newblock {\em Physical review}, 85(2):166--193, 1952.

\bibitem{bell1966problem}
John~S Bell.
\newblock On the problem of hidden variables in quantum mechanics.
\newblock {\em Reviews of Modern Physics}, 38(3):447, 1966.

\bibitem{mermin2018homer}
N~David Mermin and R{\"u}diger Schack.
\newblock {Homer} nodded: {von Neumann}’s surprising oversight.
\newblock {\em Foundations of Physics}, 48(9):1007--1020, 2018.

\bibitem{einstein1935can}
Albert Einstein, Boris Podolsky, and Nathan Rosen.
\newblock Can quantum-mechanical description of physical reality be considered complete?
\newblock {\em Physical review}, 47(10):777, 1935.

\bibitem{Whittaker_1943}
E~T Whittaker.
\newblock Chance, freewill and necessity in the scientific conception of the universe.
\newblock {\em Proceedings of the Physical Society}, 55(6):459–471, nov 1943.

\bibitem{bose1944classical}
Satyendra~Nath Bose.
\newblock The classical determinism and the quantum theory.
\newblock {\em Supplement to Current Science}, 13(1):25--27, 1944.

\bibitem{naturePiaggio1944Collapse}
H.~T.~H. Piaggio.
\newblock Collapse of determinism.
\newblock {\em Nature}, 154:122, July 1944.

\bibitem{natureWest1944Collapse}
Gilbert~D. West.
\newblock Collapse of determinism.
\newblock {\em Nature}, 154:464--465, October 1944.

\bibitem{naturePiaggio1944CollapseReply}
H.~T.~H. Piaggio.
\newblock Collapse of determinism.
\newblock {\em Nature}, 154:465, October 1944.

\bibitem{natureBarkas1944Collapse}
Wilfred~W. Barkas.
\newblock Collapse of determinism.
\newblock {\em Nature}, 154:676, November 1944.

\bibitem{natureWhittaker1944Collapse}
E.~T. Whittaker.
\newblock Collapse of determinism.
\newblock {\em Nature}, 154:676, November 1944.

\bibitem{natureGold1945Causality}
E.~Gold.
\newblock Causality or indeterminism?
\newblock {\em Nature}, 155:111, January 1945.

\bibitem{natureHorzelski1945Causality}
J.~Horzelski.
\newblock Causality or indeterminism?
\newblock {\em Nature}, 155:111, January 1945.

\bibitem{natureWest1945Causality}
Gilbert~D. West.
\newblock Causality or indeterminism?
\newblock {\em Nature}, 155:111, January 1945.

\bibitem{naturePiaggio1945Causality}
H.~T.~H. Piaggio.
\newblock Causality or indeterminism?
\newblock {\em Nature}, 155:289--290, March 1945.

\bibitem{natureBrabazon1945Causality}
{Lord Brabazon of Tara}.
\newblock Causality or indeterminism?
\newblock {\em Nature}, 155:398, March 1945.

\bibitem{pelzer1944interpretation}
H~Pelzer.
\newblock Is an interpretation of quantum phenomena by hidden parameters impossible?
\newblock {\em Proceedings of the Physical Society}, 56(3):195--197, 1944.

\bibitem{schrodinger1944collapse}
Erwin Schr{\"o}dinger.
\newblock Collapse of determinism.
\newblock Unpublished manuscript. Erwin Schr{\"o}dinger Archive, Austrian Central Library for Physics, University of Vienna, shelfmark W33-781/9, after 1944.

\bibitem{natureChanceFreewillNecessity1943}
Chance, freewill and necessity.
\newblock {\em Nature}, 151:612--613, May 1943.

\bibitem{del2026schrodinger}
Flavio Del~Santo and Nicolas Gisin.
\newblock Was {Schr{\"o}dinger} ever a determinist?
\newblock {\em Nature Reviews Physics}, pages 472 -- 473, 2026.

\bibitem{bell1964einstein}
John~S Bell.
\newblock On the {Einstein} {Podolsky} {Rosen} paradox.
\newblock {\em Physics Physique Fizika}, 1(3):195–200, 1964.

\bibitem{kaiser2020tackling}
David~I Kaiser.
\newblock Tackling loopholes in experimental tests of {Bell}'s inequality.
\newblock {\em arXiv preprint arXiv:2011.09296}, 2020.

\bibitem{DelSanto2019Realism}
Flavio Del~Santo.
\newblock Striving for realism, not for determinism: Historical misconceptions on {Einstein} and {Bohm}.
\newblock {\em APS News}, 28(5):8, May 2019.
\newblock The Back Page.

\bibitem{DelSantoKrizek2025Bohm}
Flavio Del~Santo and Gerd~Christian Krizek.
\newblock Against the `nightmare of a mechanically determined universe': Why {Bohm} was never a {Bohmian}.
\newblock In Andrea Oldofredi, editor, {\em Guiding Waves in Quantum Mechanics: One Hundred Years of de Broglie--Bohm Pilot-Wave Theory}, pages 151--170. Oxford University Press, 2025.

\bibitem{VanStrien2025Bohm}
Marij van Strien.
\newblock Why {Bohm} was never a determinist.
\newblock In Andrea Oldofredi, editor, {\em Guiding Waves in Quantum Mechanics: One Hundred Years of de Broglie--Bohm Pilot-Wave Theory}, pages 134--150. Oxford University Press, 2025.

\bibitem{hanle1979indeterminacy}
Paul~A Hanle.
\newblock Indeterminacy before {Heisenberg}: The case of {Franz Exner} and {Erwin Schr{\"o}dinger}.
\newblock {\em Historical Studies in the Physical Sciences}, 10:225--269, 1979.

\bibitem{stoltzner1999vienna}
Michael St{\"o}ltzner.
\newblock Vienna indeterminism: Mach, boltzmann, exner.
\newblock {\em Synthese}, 119(1):85--111, 1999.

\bibitem{Stoeltzner2012Schrodinger}
Michael St{\"o}ltzner.
\newblock {Erwin Schr{\"o}dinger}, {Vienna} indeterminist.
\newblock In Dennis Dieks, Wenceslao~J. Gonzalez, Stephan Hartmann, Michael St{\"o}ltzner, and Marcel Weber, editors, {\em Probabilities, Laws, and Structures}, volume~3 of {\em The Philosophy of Science in a European Perspective}, pages 481--495. Springer, Dordrecht, 2012.

\end{thebibliography}

\clearpage
\onecolumngrid

\appendix
\section{Full transcription of Schr\"odinger's unpublished manuscript ``Collapse of Determinism''}
\label{appendix}
I reproduce here a typed transcription of Schr\"odinger's handwritten unpublished manuscript \cite{schrodinger1944collapse}. Authorial deletions are omitted, while clear interlinear revisions are incorporated into the text.

\begin{center}
     \noindent\rule{3cm}{0.4pt}
\end{center}
 \vspace{0.5cm}

 \begin{center}

     I
 \end{center}

\begin{center}
     Collapse of Determinism\begingroup
\renewcommand{\thefootnote}{1)}
\footnote{Nature \underline{154}, 464; 676, 1944.}
\endgroup
\end{center}

In the second letter under this heading$^{1)}$ 
Mr. Barkas declares himself puzzled by an inference of Prof. Whittaker's
which in the latter purports to deduce from very simple facts
that in a beam of plane-polarized light which is passed through a Nicol, set at
an angle $\varphi$, the question whether a single photon is transmitted or reflected
cannot possibly be causally determined by the value of a hidden parameter in
that photon.

Prof. W.'s inference runs about thus: he assumes \underline{for simplicity} that the supposed
hidden parameter is capable only of two discrete values, say 0 and 1. Then one
would have to assume that out of $N$ photons $N\cos^2\varphi$ have the parameter
at 0; $N\sin^2\varphi$ at 1; but that sample is absurd, because the beam of light
cannot know at what angle the Nicol will be set.

I too have been puzzled by this conclusion. For the derived absurdity is visibly
produced by the simplifying assumption of two discrete parameter values only.
If you assume a parameter $x$, capable of all values between 0 and 1, which
decides for transmission whenever
\begin{equation*}
    x \geq \sin^2\varphi,
\end{equation*}
and if you assume that $N\,dx$ of $N$ photons have their $x$ between $x$ and
$x+dx$, then

 \begin{center}
     II
 \end{center}

\noindent of the $N$ 
\begin{equation*}
    \int_{\sin^2\varphi}^{1} N\,dx
    =
    N\cos^2\varphi
\end{equation*}
will be transmitted, the others reflected, as required.

For consistency one has to add the assumption that when a photon is transmitted,
its $x$ changes to
\begin{equation*}
    x' = \frac{x-\sin^2\varphi}{\cos^2\varphi},
\end{equation*}
while on reflection it changes to
\begin{equation*}
    x'' = \frac{x}{\sin^2\varphi},
\end{equation*}
in order that both the transmitted and the reflected beam have the same
composition with regard to the $x$-values of their photons as assumed for the
primary beam. I can see no inconsistency in assuming these changes. ---

Generally speaking I fully agree with Mr. Piaggio's view, that even very elaborate
and mathematically rigorous deductions, such as Von Neumann's, are of limited
value. Experimental evidence + logical deduction can \underline{never} prove that either
the deterministic or the indeterministic view is \underline{impossible}. At the very best
the sum-total of evidence can lead to the conclusion that one or the other is
--- in the sense coined by H. Poincar\'e in the case of geometry --- extremely
inconvenient, because it introduces unwarrantable hypotheses.

The strongest plea in favour of indeterminism is older than quantum mechanics. It runs about

 \begin{center}
     III.
 \end{center}
thus: we know to day [\textit{sic}] that our idea of a lawful sequence of events in Nature has
been abstracted from events which are based directly on statistical law. But from
the time when we did not know that yet (up to the 1870's) the idea of ``law of
Nature'' has become so impressed on us, that we are tempted to construct in our
mind ``true microscopic lawfulness'' behind the observed statistical one. If
that ever should prove to be unnecessary, it would be an advantage, because it
would unify our ``Weltbild''.

\end{document}